**Research Paper**

# Statistically distinguishable rating scale

## Pomazanov[1] Mikhail

[1] National Research University Higher School of Economics, Myasnitskaya St. 20, Moscow, 101000 Russian Federation; email: mhubble@yandex.ru

## ABSTRACT

The article proposes a method of designing a statistically distinguishable rating scale that is not excessive in relation to the existing observation statistics. This allows for more stable validation with a fixed maximum number of violations of the Wald criterion compared to an excess scale, which is usually used by banks. The increased robustness of validation will reduce the calibration probability of default, which provides savings in capital requirements under the advanced IRB approach. Theoretical justifications of the effect are presented, numerical calculations for three rating scales, two of which are open statistics of rating agencies, and the third is the rating scale of one bank from closed data are performed. The proposed method is most relevant for the corporate segment of the loan portfolio.

**Research Paper**

# Статистически различимая рейтинговая шкала

**Pomazanov[1] Mikhail**

[1] National Research University Higher School of Economics, Myasnitskaya St. 20, Moscow, 101000 Russian Federation; email: mhubble@yandex.ru

## ABSTRACT

В статье предложен метод проектирования статистически различимой рейтинговой шкалы, которая не является избыточной по отношению к имеющейся статистике наблюдений. Это позволяет обеспечить более устойчивую валидацию при фиксированном максимальном количестве нарушений критерия Вальда по сравнению с избыточной шкалой, которая обычно применяется банками. Повышенная устойчивость валидации позволит снизить калибровочную среднюю вероятность дефолта, что обеспечивает экономию требований к капиталу в рамках продвинутого подхода IRB. Представлены теоретические обоснования эффекта, проведены численные расчеты для трех рейтинговых шкал, две из которых – это открытые статистики рейтинговых агентств, третья - шкала одного банка из закрытых данных. Предложенный метод наиболее актуален для корпоративного сегмента кредитного портфеля.

# 1  INTRODUCTION

Существует два взаимоисключающих способа построения рейтинговой шкалы (РШ) кредитного риска на внутренних данных. Первым способом отображения (Tasche, D., 2008)[1] является условие постоянства вероятности дефолта PD во времени в разрядах РШ. Такой маппинг опирается на прямой метод, при котором сам рейтинговый балл может быть принят за PD заемщика, поскольку предполагает наличие калибровочной модели (например, статистические модели Logit, Probit , Hazard Rate).

Если в банке применяется валидированная калибровочная статистическая модель, для которой задан (текущий или прогноз) калибровочный параметр центральной тенденции PD (ЦТ) всего модельного сегмента, то после оценки вероятности дефолта заемщика его рейтинг автоматически определиться в заранее заданных внутренней РШ границах рейтинговых разрядов ( например, A2[0.5:1%), A3[1:2%), B1[2:4%) и т.д.).

$$PD(R) = \frac{1}{1+e^{\alpha R + \beta}} \to \begin{bmatrix} \dots \\ A2 = [0.5\%: 1\%) \\ A3 = [1\%: 2\%) \\ \dots \end{bmatrix}$$

Для оценки параметров $\alpha, \beta$ калибровочной функции достаточно параметров распределения рейтингового балла $R$, дискриминирующей мощности (ROC) модели и ЦТ, см., например (Pomazanov & Berezhnoy, 2024). Которые оцениваются для всего модельного сегмента с достаточной точностью даже при ограниченной статистике, причем сама калибровочная функция по определению монотонна, а значит, снижение балла означает увеличение PD.

Далее, вероятность дефолта определенного рейтингового разряда рассчитывается как среднее значение оценок вероятности дефолта отдельных заемщиков, принадлежащих каждому разряду (Basel Committee on Banking Supervision, 2023a)[2]. Для прямого метода построения РШ границы вероятности дефолта рейтингового разряда постоянны и не зависят от центральной долгосрочной тенденции PD всего модельного сегмента. Изменение со временем центральной тенденции, при неизменных показателях кредитоспособности заемщика, может привести к миграции заемщика в соседний рейтинговый разряд, после проведения периодической перекалибровки статистической модели в рамках TTC (долгосрочная) или PIT (краткосрочная) - философии. Такая миграция произойдет при выходе собственной вероятности дефолта за постоянные границы PD того разряда, в котором он был.

---

[1] стр. 177-179
[2] CRE 36.78(3)

Рейтинговый балл (или другая комплексная характеристика кредитоспособности) может и не быть однозначен PD, тогда используются косвенный метод, для которого оценка PD каждого рейтингового разряда выполняется на основе исторических показателей дефолта в разряде. Для косвенного метода неизменными являются границы рейтингового балла (или иной консолидированной характеристики кредитоспособности), но при изменении ЦТ уровня дефолтов вероятность дефолта в каждом разряде также будет изменяться. При этом, границ вероятности дефолта в разряде не существует, существуют только неизменные границы (правила) определения рейтинга по характеристикам кредитоспособности. В качестве примера в Таб.1 представлена таблица рейтинговых границ Moody's для автомобильной отрасли

**ТАБЛИЦА 1.** Пример рейтинговых границ для схемы рейтингования автомобильной отрасли (источник: Rating Methodology. Moody's Global Corporate Finance - Global Automobile Manufacturer Industry)

| Category - Sub-Category | Measurements | Weighting | Aaa | Aa | … | B | Caa |
|---|---|---|---|---|---|---|---|
| **1) Market Position and Trend** | | 35.0% | | | | | |
| a) Unit Share | Trend in global unit share over 3 years | 2.5% | >3% | >1 <3% | … | >-3 <-2% | <-3% |
| | Trend in Top 2 Market unit share over 3 years | 2.5% | >3% | >1 <3% | … | >-3 <-2% | <-3% |
| b) Product Portfolio | Product breadth and strength | 30.0% | | | Expert | | |
| **2) Leverage and Liquidity** | | 20.0% | | | | | |
| a) Leverage | 3 Yr Avg Debt / EBITDA | 5.0% | <1.00x | >1.00<2.00x | … | >5.0<6.0x | >6.0x |
| | 3 Yr Avg Debt / Capital | 5.0% | <30 % | >30< 40 % | … | >70< 80 % | >80% |
| b) Liquidity | 3 Yr Avg (Cash + Mkt Sec) / Total Debt | 10.0% | >60 % | >50 <60% | … | > 10 <20% | <10% |
| **3) Profitability and Returns** | | 15.0% | | | | | |
| a) Margin | 3 Yr Avg EBITA Margin | 7.5% | >9% | >7 <9% | … | > 1 <2% | <1% |
| b) Return | 3 Yr Avg EBITA / Average Assets | 7.5% | >10% | >8 <10% | … | > 0 <2% | <0% |
| **4) Cash Flow and Debt Service** | | 25.0% | | | | | |
| a) Cash Flow | 3 Yr Avg FCF / Debt | 5.0% | >20% | >10 <20 % | … | >-5<-2.5% | <-5% |
| | 3 Yr Avg RCF / Debt | 5.0% | >50% | >40 <50% | … | >0 <10% | <0 % |
| | 3 Yr Avg RCF / Net Debt | 5.0% | >90% | >70 <90% | … | >10 <20% | <10% |
| b) Interest Coverage | 3 Yr Avg EBIT / Interest Expense | 10.0% | >10x | >8 <10x | … | > 1 <2x | <1x |
| **5) Captive Finance Company** | | 5.0% | | | | | |
| Captive finance company | | 5.0% | | | Expert | | |
| **Total** | | 100.0% | | | | | |

Для каждого значимого фактора устанавливается рейтинговый разряд в определенных методологией границах, разряды консолидируются с установленными весами, на выходе – итоговый дискретный рейтинговый разряд. При определении рейтингового разряда для косвенного подхода, на этом этапе никакой связи с PD нет. Эта связь появляется только на следующем этапе – оценки наблюдаемой частоты дефолта для каждого рейтингового разряда.

Историческую статистику дефолтов в разряде в рамках косвенного подхода можно рассмотреть в динамике на Рис.1, Рис.2, (Moody's, 2017).

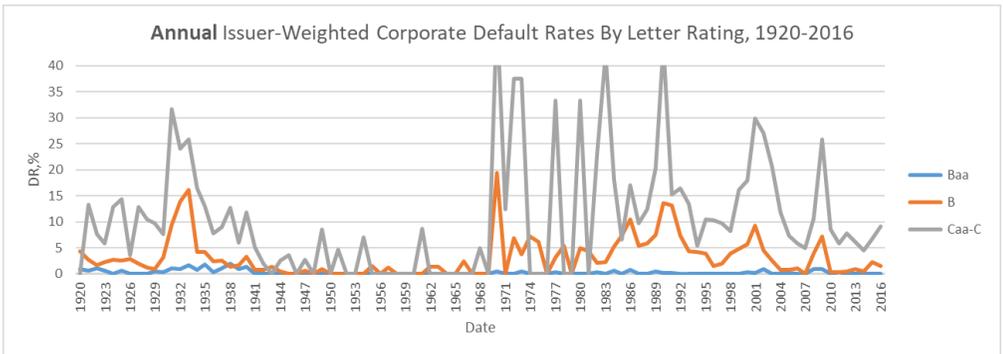

**РИСУНОК 1.** PIT (1 год) статистика дефолтов в разрядах Moody's

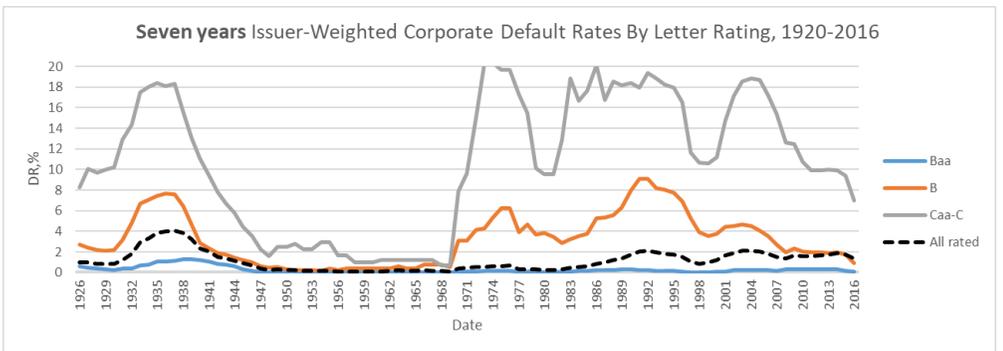

**РИСУНОК 2.** TTC (7 лет) статистика дефолтов в разрядах Moody's

Из примеров Рис.1,2 видно, что даже для TTC-частоты дефолтов в разрядах меняются существенно и никакие границы вероятностей установить постоянными невозможно.

Косвенный подход, как правило, используют крупные рейтинговые агентства, имеющие долгосрочную и достаточную статистику, прямой подход – чаще используют банки.

Косвенный подход не устанавливает заранее никакую параметрическую калибровочную монотонную функцию, а использует для калибровки PD непараметрический подход наблюдений за частотой дефолтов в каждом разряде. Однако, здесь заложен самый значимый на практике недостаток такого подхода, поскольку даже самая эффективная рейтинговая модель не гарантирует монотонности PD по разрядам при ограниченной статистике наблюдений. Например, текущая статистика транснационального рейтингового агентства Fitch (Fitch Ratings, 2024)[3] не дает монотонность частот дефолтов по рейтинговым разрядам за 34 года наблюдении (Таб.2.)

**ТАБЛИЦА 2**. Средние глобальные годовые частоты дефолтов по корпоративным финансам по версии Fitch: 1990 - 2023

| Рейтинг | Среднегодовое количество наблюдений | Частота дефолтов, %, среднегодовая | Монотонность |
|---|---|---|---|
| AAA | 922 | 0,108 | - |
| AA+ | 643 | 0,000 | 🟥 |
| AA | 1 780 | 0,000 | 🟥 |
| AA- | 4 028 | 0,074 | 🟩 |
| A+ | 5 604 | 0,000 | 🟥 |
| A | 8 005 | 0,062 | 🟩 |
| A- | 7 683 | 0,065 | 🟩 |
| BBB+ | 8 497 | 0,082 | 🟩 |
| BBB | 9 096 | 0,066 | 🟥 |
| BBB- | 7 869 | 0,216 | 🟩 |
| BB+ | 3 862 | 0,259 | 🟩 |
| BB | 3 573 | 0,504 | 🟩 |
| BB- | 3 547 | 1,071 | 🟩 |
| B+ | 3 109 | 1,447 | 🟩 |
| B | 3 193 | 1,942 | 🟩 |
| B- | 2 104 | 3,042 | 🟩 |
| CCC to C | 1 300 | 23,308 | 🟩 |
| **Aggregated & weighted average** | **74 815** | **0,781** | |

Объяснение этому простое – статистические флуктуации, поскольку отсутствуют причины сомневаться в корректности методологии Fitch, чтобы заподозрить методику рейтингования в неполной адекватности. Если, при внедрении косвенного подхода в банке, будет наблюдаться подобная немонотонность, то для экономии стоимости риска будет выгодно манипулятивно понижать рейтинг для снижения резервов и капитала. Это формально не противоречит консервативности. Например, если у компании расчетный рейтинг оказался BBB+ с вероятностью дефолта 0.082%, то выгодно его понизить до BBB, получив меньшую вероятность 0.066%, согласно Таб. 2. Однако, фактически шкала разрушается при таких манипуляциях.

---

[3] Fitch Global Corporate Finance Transition Matrices: 1990–2023

Для решения проблемы монотонности РШ применяются сглаживающие алгоритмы. Например, (The Credit Research Initiative of the National University of Singapore, 2017) использует линейную регрессию logit-преобразования частоты дефолтов в разрядах. Минусом такого подхода является неучет статистики наблюдений в разрядах, а также неучет нулевых значений частот для тех разрядов, где дефолтов не наблюдается. Другим недостатком такого простого сглаживания (с точки зрения требований к банкам), является неучет ограничения минимальной вероятности дефолта в наивысшем разряде РШ банка, которое составляет 0.05% (Basel Committee on Banking Supervision, 2023c)[4]. Таких недостатков лишен подход к сглаживанию, предложенный в работе (Huajian, 2018), который основан на методе максимального правдоподобия с параметрами, количество которых равно количеству разрядов. Ниже будет раскрыта суть метода в конечных формулах, которые будут использоваться в предложенных комплексах алгоритмов, являющиеся целью настоящей работы.

## 2 КОНЦЕПЦИЯ СТАТИСТИЧЕСКОЙ РАЗЛИЧИМОСТИ РЕЙТИНГОВОЙ ШКАЛЫ

### 2.1 Проблема неоднозначности требований к количеству разрядов рейтинговой шкалы

Базельский комитет и Европейское банковское управление выставляют достаточно общие регуляторные требования к количеству разрядов РШ (Basel Committee on Banking Supervision, 2023a)[5], (EBA, 2023)[6]:

– РШ заемщиков должна содержать не менее 8 разрядов, из которых не менее 7 разрядов для заемщиков, не находящихся в состоянии дефолта, и 1 разряд для заемщиков, находящихся в состоянии дефолта;
– РШ должна иметь достаточное количество разрядов, которое позволяет исключить наличие высокой концентрации заемщиков в определенных разрядах РШ;

---

[4] CRE 32.4
[5] CRE 36.19
[6] 36.c

– разряд РШ должен охватывать диапазоны величины вероятности дефолта, которые позволяют исключить наличие высокой концентрации заемщиков, отнесенных к одному разряду.

Ограничение сверху на допустимое количество разрядов отсутствует, Регулятор ничего не требует вплоть до абсурдного, когда разрядов больше, чем наблюдений или половина – пустые. Вместе с тем регулятор предполагает, что основной статистический тест сопоставительного анализа соответствия средней вероятности дефолта в разряде наблюдаемой частоте дефолта и формирующий контрольный показатель качества РШ - является критерий Вальда биномиального теста на консервативность (Basel Committee on Banking Supervision, 2005)[7]. Биномиальный тест с критерием Вальда проводится для каждого разряда РШ для уровней значимости $\alpha = 5\%$ и $\alpha = 1\%$ и требует[8], чтобы

$$PD_i + \Phi^{-1}(1-\alpha)\sqrt{\frac{PD_i(1-PD_i)}{n_i}} > DR_i, \qquad (2.1)$$

где $n_i$ - количество наблюдений в i-ом разряде РШ; $PD_i$ - среднее значение вероятности дефолта заемщиков в i-ом разряде РШ, используемое при расчете величины кредитного риска; $DR_i$ - значение уровня дефолта в i-ом разряде РШ; $\Phi^{-1}(x)$ - обратная функция стандартного нормального распределения. Причем, если критерий не выполняется при $\alpha = 1\%$, то разряд РШ признается значительно недооцененным (красная зона валидации), а если выполняется при $\alpha = 5\%$, то разряд РШ признается корректно оцененным (зеленая зона валидации). Шкала в целом оценивается по количеству «желтых» и красных разрядов, например, если оказалось 5 и более «желтых» разрядов, то РШ признается неудовлетворительной (т.е. красной»).

На практике многие банки применяют избыточное количество разрядов РШ для корпоративных кредитных портфелей, где нет большой размерности наблюдений, поскольку отсутствуют ограничения. Аргументируя такое решение разными причинами, такими как: синхронизировать количество разрядов с рейтинговыми агентствами, увеличить «точность» принятия кредитного решения, уменьшить ошибки округления PD негативно влияющие на требования к капиталу и т.п. Существует ли фундаментальный ограничитель? Оказывается, да.

У каждого разряда РШ существует среднее PD, обозначим $p^*$; а также нижняя и верхняя граница PD разряда[9], обозначим $p^L, p^U$. Доверительный интервал попадания $DR$ в разряд при заданном уровне достоверности $\alpha$ не должен выходить за пределы отрезка $p \in [p^L, p^U]$, в противном случае разряды становятся статистически неразличимы в рамках заданного количества возможных наблюдений $N = \sum_i n_i$. В рамках критерия Вальда нетрудно установить минимальное требование количества наблюдений $m_\alpha$, обеспечивающее различимость отрезка $[p^L, p^U]$ на уровне достоверности $\alpha$:

$$m_\alpha = \left\lceil \frac{z_{\alpha/2}^2(1-p^*)}{\epsilon_R^2 p^*} \right\rceil, \qquad (2.2)$$

где $\epsilon_R$ - относительный допуск отклонения от средней вероятности дефолта в разряде, $\epsilon_R = min\left(\frac{p^*}{p^L}, \frac{p^U}{p^*}\right) - 1$; $z_{\alpha/2} = \Phi^{-1}\left(1 - \frac{\alpha}{2}\right)$; знак $\lceil \cdot \rceil$ - округление до верхнего целого.

---

[7] PP. 47

[8] Формулой (2.1) задан асимптотический критерий, достаточный для практики при условии $\min(PD \cdot N, (1-PD) \cdot N) > 10$. Точный критерий Вальда
$\frac{1}{N} min \left\{ k \left| \sum_{n=k}^{N} \binom{N}{n} PD^n (1-PD)^{N-n} \leq 1-\alpha \right. \right\} > DR$.

Проблема минимального количества наблюдений для внедрения статистических критериев с заданным уровнем достоверности не нова и многократно рассматривалась в научной литературе. Формула (2.2) взята из работы (McGrath & Burke, 2024)[10], в которой авторы рассмотрели также и более тонкие критерии биномиального теста.

Используя аналогичную логику, которой пользуются при признании зоны разряда по критерию Вальда, справедливы следующие рассуждения: если количество наблюдений в разряде $n < m_{5\%}$, значит мы на том же уровне значимости не сможем различить «красный» (значительно недооцененный) разряд и вынуждены разряд признать «серым». Поскольку критерий (2.2) будет допускать выход в соседний разряд за пределы отрезка $[p^L, p^U]$ при недостатке наблюдений. Если $m_{5\%} \leq n < m_{1\%}$, то возможно отличие «желтого» (умеренная недооценка) от «зеленого» и только если количество наблюдений достаточно $n \geq m_{1\%}$, то это полноценный («цветной») разряд для валидации по критерию Вальда (2.1).

## 2.2 Статистически различимая рейтинговая шкала

РШ признается статистически различимой на уровне значимости $\alpha$, если для всех разрядов имеется достаточное количество наблюдений, превышающее $m_\alpha$ (2.2).

Для практики достаточно брать $\alpha = 5\%$.

Проведем проектирование на основе формулы (2.2). Пусть известен риск-профиль субъектов наблюдений проектируемой различимой РШ, заданный их плотностью $f(p)$ по аргументу вероятности дефолта, $\int_0^1 f(x)dx = 1$, $F(p) = \int_0^p f(x)dx$. Обозначим нижнюю и верхнюю границу PD разряда $p^L = p, p^U = \lambda p$, где параметр $\lambda > 1$, размерность $N$ потенциальных совокупных наблюдений считаем заданной.

---

[9] В рамках прямого подхода к формированию РШ - эти границы закреплены нормативно, в рамках косвенного подхода – определяются тем или иным способом.

[10] формула (3)

Тогда средняя вероятность дефолта в разряде, за исключением первого, будет определяться по формуле

$$p^*(p,\lambda) = \frac{\int_p^{\lambda p} xf(x)dx}{F(\lambda p) - F(p)} = \frac{p \cdot (\lambda F(\lambda p) - F(p)) - \int_p^{\lambda p} F(x)dx}{F(\lambda p) - F(p)}.$$

Первый разряд с верхней границей $p^U = p$ будет определять среднюю PD как

$p_1^*(p) = \frac{pF(p) - \int_0^p F(x)dx}{F(p)}$. Из формулы (2.2) рекуррентным образом определяются границы $p_i^L, p_i^U$, при этом $p_{i+1}^L = p_i^U, p_1^L = 0, p_G^U = 1$, где $G$ - количество грейдов.

Уравнения для проектирования различимой РШ определяются одним из двух возможных неявных каскадов, в каждом из которых необходимо последовательно решать уравнения.

**Восходящий каскад** (от низко-дефолтного разряда РШ к высоко-дефолтному)

$$\left[ \begin{array}{l} i = 1, N = \left\lceil \dfrac{z_{\alpha/2}^2(1-p_1^*(x))}{\left(\dfrac{x}{p_1^*(x)}-1\right)^2 \cdot p_1^*(x) \cdot F(x)} \right\rceil \Rightarrow x \Rightarrow p_1^U = x \\ i \geq 2, N = \left\lceil \dfrac{z_{\alpha/2}^2(1-p^*(p_i^U,\lambda))}{\left(min\left(\dfrac{p^*(p_i^U,\lambda)}{p_i^U}, \dfrac{\lambda p_i^U}{p^*(p_i^U,\lambda)}\right)-1\right)^2 \cdot p^*(p_i^U,\lambda) \cdot \left(F(\lambda p_i^U) - F(p_i^U)\right)} \right\rceil \Rightarrow \lambda \Rightarrow \begin{array}{l} p_{i+1}^U = \lambda p_i^U \\ p_{i+1}^L = p_i^U \end{array} \end{array} \right. \quad (2.3)$$

Рекуррентный каскад (2.3) останавливается, если $p_{i+1}^U \geq 1$, тогда количество не дефолтных разрядов $G = i+1, p_{i+1}^U = 1$ либо уравнение (2.3) не имеет решения, тогда $G = i, p_i^U = 1$. Второй вариант остановки (2.3) реализуется, когда оставшихся на последнем шаге наблюдений не хватает для обеспечения минимального количества наблюдений для различимости последнего (низшего) разряда, определяемого (2.2).

**Нисходящий каскад** (от высоко-дефолтного разряда РШ к низко-дефолтному)

Очевидно, $p_1^U = 100\%$, обозначим

$$p^\#(p,\lambda) = \frac{\int_{p/\lambda}^p xf(x)dx}{F(p) - F(p/\lambda)} = \frac{p \cdot \left(F(p) - \frac{F(p/\lambda)}{\lambda}\right) - \int_{p/\lambda}^p F(x)dx}{F(p) - F(p/\lambda)}, \quad p_1^\#(p) = \frac{1 - p \cdot F(p) - \int_p^1 F(x)dx}{1 - F(p)}$$

$$\left[ \begin{array}{l} i = 1, N = \left\lceil \dfrac{z_{\alpha/2}^2(1-p_1^\#(x))}{\left(min\left(\dfrac{p_1^\#(x)}{x}, \dfrac{1}{p_1^\#(x)}\right)-1\right)^2 \cdot p_1^\#(x) \cdot (1-F(x))} \right\rceil \Rightarrow x \Rightarrow p_1^L = x, p_2^U = p_1^L \\ i \geq 2, N = \left\lceil \dfrac{z_{\alpha/2}^2(1-p^\#(p_i^U,\lambda))}{\left(min\left(\dfrac{p^\#(p_i^U,\lambda) \cdot \lambda}{p_i^U}, \dfrac{p_i^U}{p^\#(p_i^U,\lambda)}\right)-1\right)^2 \cdot p^\#(p_i^U,\lambda) \cdot \left(F(p_i^U) - F(p_i^U/\lambda)\right)} \right\rceil \Rightarrow \lambda \Rightarrow \begin{array}{l} p_{i+1}^U = \dfrac{p_i^U}{\lambda} \\ p_i^L = p_{i+1}^U \end{array} \end{array} \right. \quad (2.4)$$

Рекуррентный каскад (2.4) останавливается, если уравнение (2.4) не имеет решения, тогда $G = i, p_i^L = 0$, т.е. оставшихся на последнем шаге наблюдений не хватает для обеспечения минимального количества наблюдений для различимости наивысшего разряда, определяемого (2.2). Это допустимо, поскольку все разряды, кроме наивысшего, различимы. Процесс (2.3) или (2.4) разделит ось PD на грейды, в каждом из которых количество наблюдений будет минимально достаточным для различимости РШ. Рекурентный процесс разрешим при строгой монотонности $p_i^L, p_i^U$. Индекс концентрации Херфиндалая-Хиршмана, который является ключевым показателем качества РШ, будет оцениваться как

$$HHI = F(p_1)^2 + \sum_{i=2}^{G} \left(F(p_i^U) - F(p_{i-1}^U)\right)^2$$
— для восходящего каскада
$$HHI = F(p_G^U)^2 + \sum_{i=2}^{G} \left(F(p_{i-1}^U) - F(p_i^U)\right)^2$$
— для нисходящего каскада
(2.5)

Скорректированный индекс концентрации $HHI_{adj} = \frac{HHI - 1/G}{1 - 1/G}$.

Итого, проектирование различимой рейтинговой шкалы проводится по гиперпараметрам:
1. Уровень значимости $\alpha$ (рекомендуется 5%);
2. Профиль риска - распределение наблюдений по PD;
3. Потенциальная размерность наблюдений, которая в перспективе имеется «в руках» у разработчика РШ, например, N=10000.

Далее применяется рекуррентный процесс (2.3) или (2.4).

## 3 ПРАКТИЧЕСКИЕ РАСЧЕТЫ НА ПРИМЕРАХ ШКАЛ РЕЙТИНГОВЫХ АГЕНТСТВ

### 3.1 Статистическая различимость сглаженных шкал рейтинговых агентств Fitch и Эксперт-РА

Для исследования статистической различимости с учетом общего количества (размерности) наблюдений, на основе которых определяется частота годовых дефолтов в каждом рейтинговом разряде, РШ мы возьмем данные двух рейтинговых агентств Fitch (участник Большой тройки) и Эксперт-РА (национальное рейтинговое агентство, Россия). Для первого рассматривается глобальные годовые частоты дефолтов по корпоративным финансам (Fitch Ratings, 2024), для второго – все исторические данные об уровнях дефолта по рейтинговым категориям (Эксперт РА, 2024).

Для применения формулы минимального требования количества наблюдений $m_\alpha$ (2.2) необходимо установить границы вероятности в каждом разряде $[p^L, p^U]$, а также значение вероятности дефолта $p^*$ так, чтобы значения таких средних были монотонны от разряда к разряду. Дополнительно, на монотонность особо обращает внимание Регулятор (смотреть, например (EBA, 2023)[11]), полагая, что не монотонность - это исключение, требующее обоснования.

Используется метод (Huajian, 2018), который состоит в следующем. Пусть в каждом не дефолтном разряде $i = 1 \ldots G$ суммарное количество годовых наблюдений $n_i > 0$, из них дефолтных $d_i \geq 0$, тогда логарифм функции правдоподобия

$$LL(p) = \sum_{i=1}^{G} d_i \cdot \ln(p_i) + (n_i - d_i) \cdot \ln(1 - p_i) \quad (3.1)$$

Очевидно, что при отсутствии ограничений максимум (3.1) достигается при не сглаженных значениях $p_i = d_i/n_i$, поэтому накладываются ограничения на $p_i$. Значения вероятностей в разрядах параметризуются в виде

$$\begin{aligned} p_i &= exp(b_i + b_2 + \cdots + b_G) \\ b_G &< 0, b_1 = \cdots = b_{G-1} = -\epsilon \leq 0 \end{aligned} \quad (3.2)$$

Далее решается задача максимума правдоподобия (3.1) относительно параметров $b_1, b_2, \ldots b_G$ при ограничениях (3.2), обеспечивающих монотонность, на выходе - $p_i^*$. Учитывается ограничение регулятора для банков $p_1 \geq 0.05\%$, т.е. $\sum_{i=1}^{G} b_i \geq \ln(0.05\%)$. Для определения граничных значений $p^L, p^U$ выбираются средние точки в последовательности $b_1, b_2, \ldots b_G$:

$$\begin{aligned} p_1^L &= 0, p_i^L = exp\left(\sum_{k=i-1}^{G} b_k - \frac{b_{i-1}}{2}\right) \\ p_i^U &= exp\left(\sum_{k=i}^{G} b_k - \frac{b_i}{2}\right), p_G^U = 1 \end{aligned} \quad (3.3)$$

Результат расчетов по сглаживанию и применению формулы (2.2) для сравнения статистик приведен в Таб. 3 и Таб. 4 для двух целевых рейтинговых агентств.

**ТАБЛИЦА 3**. Различимость шкалы рейтингового агентства Fitch: 1990 - 2023 (параметр моностонности $\epsilon = 0.1$)

| Rating | DR | $p^L$ | $p^*$ | $p^U$ | $n_i$ | $m_{5\%}$ | Distinguish-ability |
|---|---|---|---|---|---|---|---|
| AAA | 0,11% | 0,000% | 0,050% | 0,053% | 31 348 | 2 921 225 | 🟥 |

---

[11] 38.d

| Rating | DR | $p^L$ | $p^*$ | $p^U$ | $n_i$ | $m_{5\%}$ | Distinguish-ability |
|---|---|---|---|---|---|---|---|
| AA+ | 0,00% | 0,053% | 0,055% | 0,058% | 21 862 | 2 643 095 | 🟥 |
| AA | 0,00% | 0,058% | 0,061% | 0,064% | 60 520 | 2 391 432 | 🟥 |
| AA– | 0,07% | 0,064% | 0,067% | 0,071% | 136 952 | 2 163 718 | 🟥 |
| A+ | 0,00% | 0,071% | 0,075% | 0,078% | 190 536 | 1 957 674 | 🟥 |
| A | 0,06% | 0,078% | 0,082% | 0,087% | 272 170 | 1 771 238 | 🟥 |
| A– | 0,07% | 0,087% | 0,091% | 0,096% | 261 222 | 1 602 543 | 🟥 |
| BBB+ | 0,08% | 0,096% | 0,101% | 0,106% | 288 898 | 1 449 902 | 🟥 |
| BBB | 0,07% | 0,106% | 0,111% | 0,150% | 309 264 | 1 311 787 | 🟥 |
| BBB- | 0,22% | 0,150% | 0,202% | 0,236% | 267 546 | 67 989 | 🟩 |
| BB+ | 0,26% | 0,236% | 0,275% | 0,369% | 131 308 | 49 880 | 🟩 |
| BB | 0,50% | 0,369% | 0,494% | 0,717% | 121 482 | 6 726 | 🟩 |
| BB– | 1,07% | 0,717% | 1,042% | 1,235% | 120 598 | 10 601 | 🟩 |
| B+ | 1,45% | 1,235% | 1,464% | 1,699% | 105 706 | 10 015 | 🟩 |
| B | 1,94% | 1,699% | 1,972% | 2,437% | 108 562 | 7 396 | 🟩 |
| B– | 3,04% | 2,437% | 3,012% | 8,393% | 71 536 | 2 226 | 🟩 |
| CCC- C | 23,31% | 8,393% | 23,393% | 100,000% | 44 200 | 4 | 🟩 |
| Total | 0,781% | | 0,799% | | 2 543 710 | 18 367 451 | |

**ТАБЛИЦА 4.** Различимость шкалы рейтингового агентства Эксперт-РА: 2001-2024 (параметр моностонности $\epsilon = 0.1$)

| Rating | DR | $p^L$ | $p^*$ | $p^U$ | $n_i$ | $m_{5\%}$ | Distinguish-ability |
|---|---|---|---|---|---|---|---|
| ruAAA | 0,00% | 0,00% | 0,05% | 0,05% | 365 | 2 921 215 | 🟥 |
| ruAA+ | 0,00% | 0,05% | 0,06% | 0,08% | 206 | 2 643 086 | 🟥 |
| ruAA | 0,25% | 0,08% | 0,11% | 0,12% | 396 | 1 281 076 | 🟥 |
| ruAA- | 0,00% | 0,12% | 0,13% | 0,26% | 372 | 1 159 026 | 🟥 |
| ruA+ | 0,61% | 0,26% | 0,54% | 0,57% | 488 | 269 962 | 🟥 |
| ruA | 0,74% | 0,57% | 0,60% | 0,63% | 538 | 244 133 | 🟥 |
| ruA- | 0,51% | 0,63% | 0,66% | 0,91% | 593 | 220 762 | 🟥 |
| ruBBB+ | 1,19% | 0,91% | 1,26% | 1,48% | 505 | 10 307 | 🟥 |
| ruBBB | 2,14% | 1,48% | 1,73% | 1,82% | 513 | 83 021 | 🟥 |
| ruBBB- | 1,61% | 1,82% | 1,91% | 2,44% | 684 | 74 981 | 🟥 |
| ruBB+ | 3,14% | 2,44% | 3,11% | 3,84% | 828 | 2 192 | 🟥 |
| ruBB | 5,06% | 3,84% | 4,74% | 4,98% | 613 | 29 387 | 🟥 |
| ruBB- | 6,89% | 4,98% | 5,24% | 5,50% | 305 | 26 452 | 🟥 |
| ruB+ | 6,27% | 5,50% | 5,79% | 6,08% | 335 | 23 796 | 🟥 |
| ruB | 5,29% | 6,08% | 6,39% | 6,72% | 359 | 21 392 | 🟥 |
| ruB- | 4,98% | 6,72% | 7,07% | 9,75% | 221 | 19 217 | 🟥 |
| ruCCC | 13,30% | 9,75% | 13,44% | 19,02% | 218 | 173 | 🟩 |
| ruCC | 28,57% | 19,02% | 26,91% | 100,00% | 21 | 61 | 🟥 |

| Total | 2,685% | 2,685% | 7560 | 6 109 024 |
|---|---|---|---|---|

Видно, что рейтинговая шкала Fitch различима для разрядов не выше BBB-, несмотря на обширную статистику. Шкала национального рейтингового агентства Эксперт-РА не различима, за исключением одного предпоследнего разряда. Что ожидаемо, поскольку статистика 7560 наблюдений небогатая для такого количества разрядов. Приведенные кейсы указывают на целесообразность построения различимой РШ.

### 3.2 Проектирование статистически различимых рейтинговых шкал для переменной размерности наблюдений

Пусть $P_1$, $P_2$ и т. д. – правые границы вероятности дефолта $p^U$ в каждом разряде $k = 1, \ldots, G$, при этом подразумевается строгая монотонность $P_i < P_{i+1}$  $n_k$ – нормированные концентрации наблюдений, $\sum_{k=1}^{G} n_k = 100\%$. Правая граница разряда $k$ равна левой границе разряда $k + 1$ и т.д. Учитывая близость распределения PD в разрядах к геометрической прогрессии, для получения релевантной функции распределения $F(p)$ в качестве аргумента положим $ln(p)$ для $p > P_1$. Кусочно-непрерывная функция распределения

$$F(p) = \begin{cases} \dfrac{n_1}{P_1} p, & p < P_1 \\ n_k \cdot \dfrac{\ln(p) - \ln(P_{k-1})}{\ln(P_k) - \ln(P_{k-1})} + \sum_{i=1}^{k-1} n_i, & P_{k-1} \leq p < P_k \\ 1, & p \geq 1 \end{cases} \quad (3.4)$$

Для строгой монотонности шкалы и исполнения формулы (3.4) необходимо в общем случае провести операцию сглаживания (3.1), (3.2), (3.3).

Проведем расчеты для трех рейтинговых шкал, два из которых – это Fitch и Эксперт-РА, а третья – это корпоративная шкала внутреннего кредитного рейтинга (ICR) одного из банков[12], которая имеет избыточную размерность $G = 21$. Сглаженные профили риска $F(PD)$ названных шкал представлены на Рис. 3.

---

[12] Источник не раскрывается

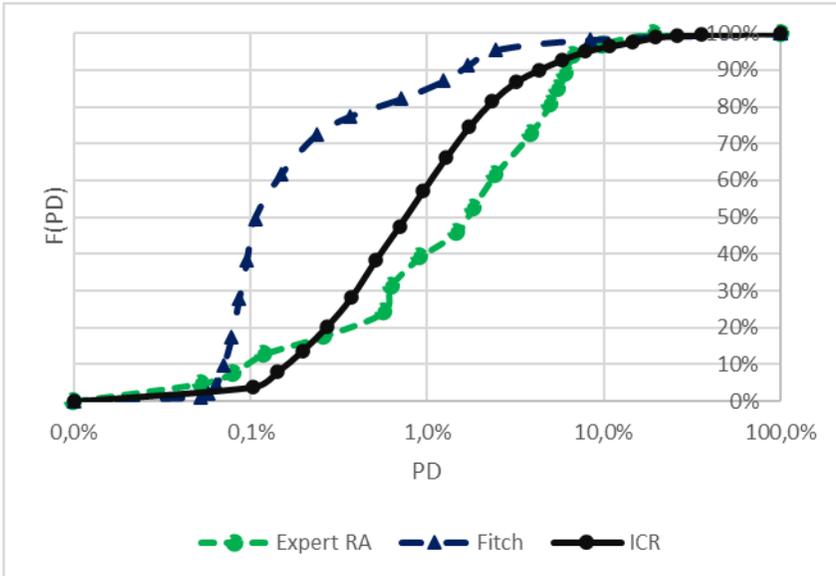

**РИСУНОК 3.** Сглаженные профили риска для трех объектов: рейтинговое агентство Эксперт-РА, рейтинговое агентство Fitch, внутренний кредитный рейтинг (ICR) одного из банков.

Проектирование статистически различимых РШ в зависимости от переменной размерности наблюдений проводится в варианте восходящего каскада (2.3). На Рис. 4 представлена зависимость размерности минимального количества статистически различимых шкал от размерности наблюдений для трех источников риск-профиля, указанных выше.

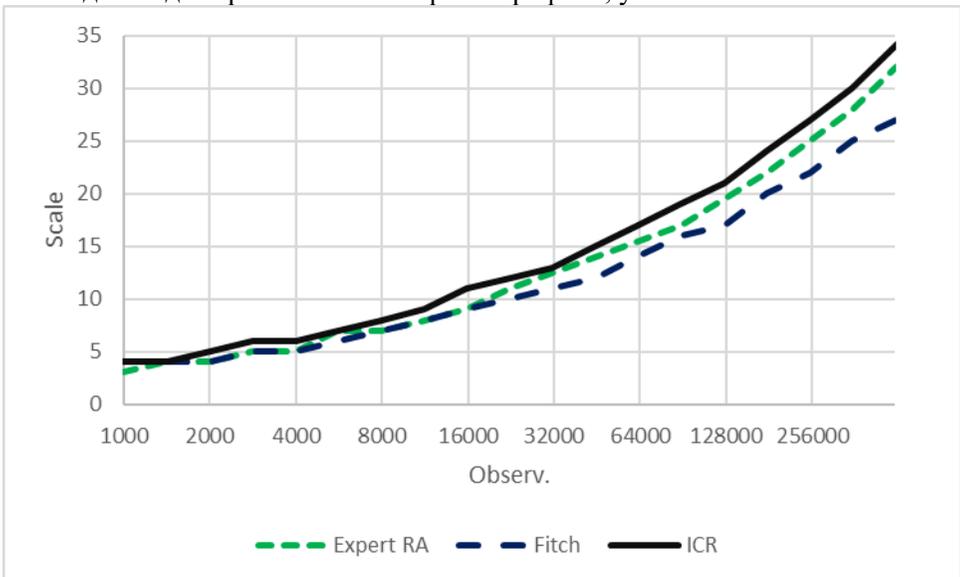

**РИСУНОК 4.** Зависимость размерности минимального количества статистически различимых шкал от размерности наблюдений для трех источников риск-профиля: Эксперт-РА, Fitch и ICR.

Верхняя граница вероятности дефолта в первом (наивысшем) разряде РШ будет зависеть от размерности всех наблюдений. На Рис. 5 представлены эти зависимости.

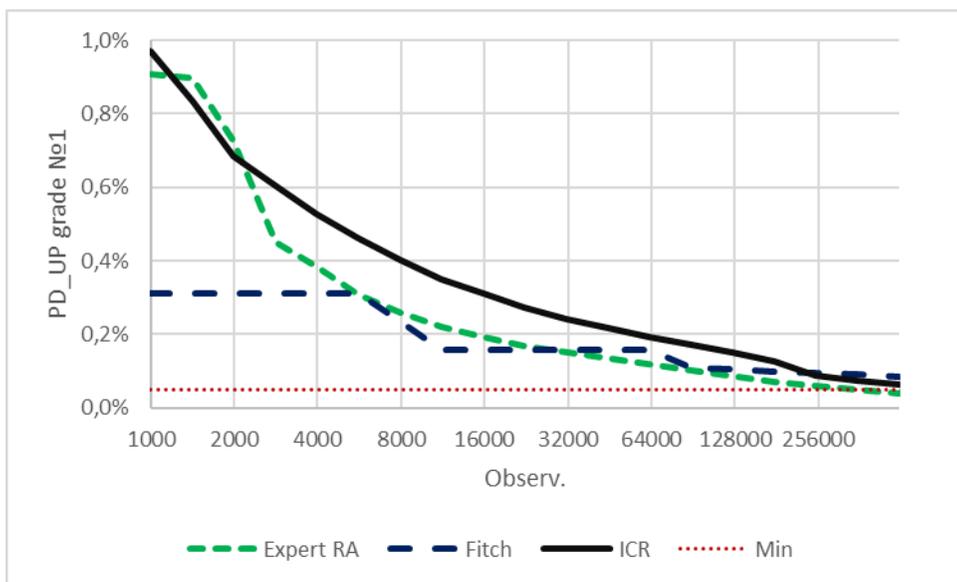

**РИСУНОК 5.** Зависимость верхней границы вероятности дефолта в наивысшем разряде в зависимости от размерности наблюдений. Источники риск-профилей: Эксперт-РА, Fitch, ICR.

Из Рис.5 видно, что риск-профиль оказывает существенное влияние на уровень дефолта в первом разряде при лимитированной размерности наблюдений. Для количества наблюдений 200000 и более все риск-профили показывают приближение значения $PD_1$ к минимальному разрешенному регулятором значению $PD_{Min} = 0.05\%$.

В силу алгоритма построения статистически различимой РШ в варианте восходящего каскада (2.3) разумно ожидать, что распределение относительного количества наблюдений по разрядам будет убывать с понижением разряда (повышением номера) и на высший разряд выпадет максимальная размерность наблюдений. На Рис. 6 представлено распределение концентраций в различимых недефолтных разрядах для 10000 наблюдений риск-профиля ICR.

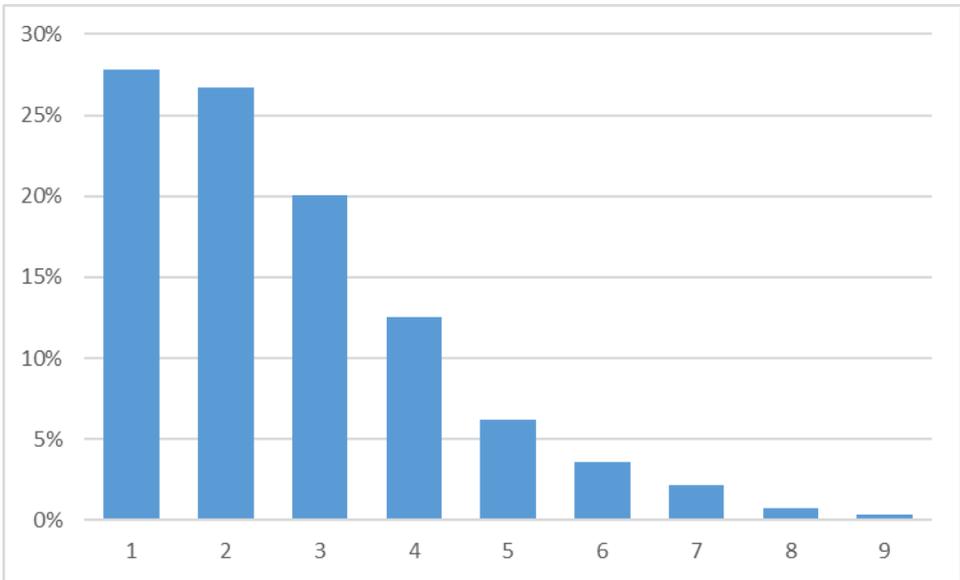

**РИСУНОК 6**. Распределение концентраций в различимых разрядах для 10000 наблюдений риск-профиля ICR.

Профиль концентраций РШ, спроектированной исходя из требования статистической различимости, снижает избыточность разрядов, что может привести к нарушению требований регулятора недопустимости высокой концентрации РШ. Метрика концентраций определена формулой (2.5) и контролируется показателем $HHI_{adj}$. На Рис. 7 представлены зависимости $HHI_{adj}$ от размерности наблюдений для трех риск-профилей, указанных выше.

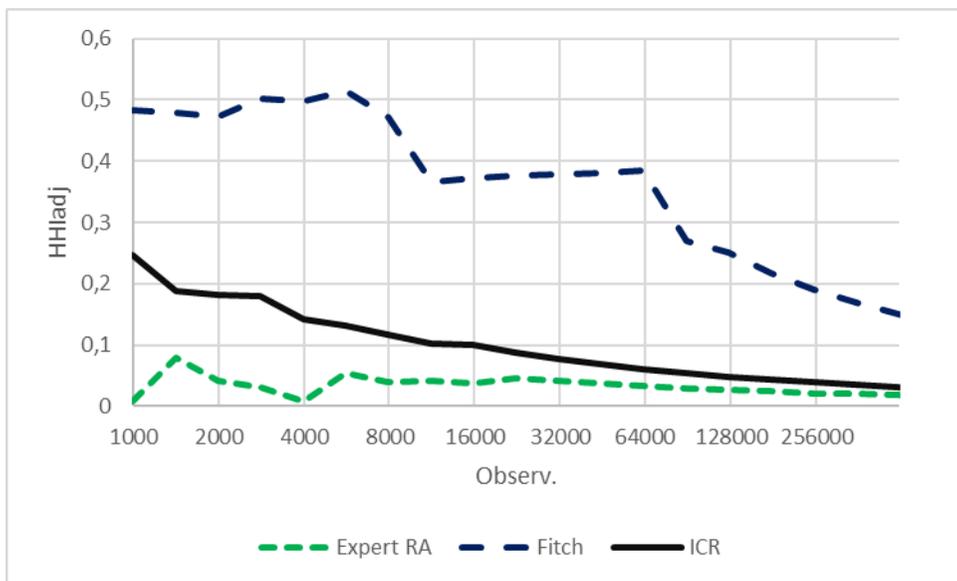

**РИСУНОК 7.** Зависимость метрики $HHI_{adj}$ концентраций РШ от количества наблюдений. Источники риск-профилей: Эксперт-РА, Fitch, ICR.

На практике применяется требование $HHI_{adj} < 0.2$. Из Рис.7. видно, что риск-профили Эксперт-РА и ICR порождают различимые шкалы удовлетворительной концентрации для размерности наблюдений от 2-4 тысяч, риск-профиль Fitch требует повышенной размерности наблюдений для удовлетворения минимальных требований к концентрации.

## 4 ЭФФЕКТИВНОСТЬ ВНЕДРЕНИЯ СТАТИСТИЧЕСКИ РАЗЛИЧИМОЙ РЕЙТИНГОВОЙ ШКАЛЫ

### 4.1 Теоретические предпосылки

Рассмотрим асимптотический критерий Вальда (2.1) биномиального теста для выборки из $n$ наблюдений, вероятность дефолта каждого равна $p$, вероятность дефолта соответствует частоте дефолта так, что для наблюдаемых $k$ дефолтов вероятность $P(k > k^*) = \alpha$, где $k^* = np + z_\alpha\sqrt{np(1-p)}$, $z_\alpha$ аргумент обратного нормального распределения $z_\alpha = \Phi^{-1}(1-\alpha)$. То есть

$$P(k > k^*) = 1 - \Phi\left(\frac{k^* - np}{\sqrt{np(1-p)}}\right) \quad (4.1)$$

Зададим граничное значение $k_\varepsilon^*$ предполагая (закладывая в калибровку), что вероятность дефолта это $p_\varepsilon = p \cdot (1 - \varepsilon)$, где $\varepsilon$ малая величина. Тогда, следуя (4.1), вероятность того, что «валидация провалится» $P(k > k_\varepsilon^*) = 1 - \Phi\left(\frac{k_\varepsilon^* - np}{\sqrt{np(1-p)}}\right)$. При этом $k_\varepsilon^* = np(1-\varepsilon) + z_\alpha\sqrt{np(1-p+p\varepsilon)(1-\varepsilon)} = k^* - \varepsilon\left(np + z_\alpha\left(1 - \frac{1}{2(1-p)}\right)\sqrt{np(1-p)}\right) + o(\varepsilon^2)$, что, после подстановки в формулу (4.1), дает

$$P(k > k_\varepsilon^*) = \alpha + \varepsilon\left(\sqrt{\frac{np}{1-p}} + z_\alpha\left(1 - \frac{1}{2(1-p)}\right)\right)\varphi(z_\alpha) + o(\varepsilon^2) \quad (4.2)$$

где $\varphi(x) = \Phi'(x)$.

Предположим, что выборка была разбита на две части, в каждой из частей вероятности дефолта $p_1, p_2$ соответствуют частоте дефолта, концентрации $n_1, n_2, n = n_1 + n_2$. Совокупная частота дефолта $p = (p_1 n_1 + p_2 n_2)/n$ тоже соответствует частоте дефолтов соответственно. Вероятность $P_V$ получить «провал валидации» хотя-бы на одной из двух выборок на уровне значимости $\alpha$ $P_V = P(\{k_1 > k_1^*\} \vee \{k_2 > k_2^*\}) = P(k_1 > k_1^*) + P(k_1 > k_1^*) - P(k_1 > k_1^*) \cdot P(k_1 > k_1^*) = 2\alpha - \alpha^2$. После объединения выборок в одну вероятность «провала валидации» будет определяться формулой (4.2), которая дает преимущество в том, что при удержании вероятности «провала валидации» на том-же уровне, что и для двух выборок, для объединённой выборки появляется возможность снизить калибровочную вероятность дефолта на относительную величину $\varepsilon$ при условии

$$\Omega = \sqrt{\frac{np}{1-p}} + z_\alpha\left(1 - \frac{1}{2(1-p)}\right) > 0 \quad (4.3)$$

Проверим условие (4.3) для статистически различимой РШ, обладающей свойством (2.2), т.е. при $n = m_\alpha$. Тогда $\Omega = \frac{z_{\alpha/2}}{\epsilon_R} + z_\alpha\left(1 - \frac{1}{2(1-p)}\right)$ и условие (4.3) превращается в условие

$$p < 1 - \frac{z_\alpha}{2\left(z_\alpha + \frac{z_{\alpha/2}}{\epsilon_R}\right)} \quad (4.4)$$

(4.4) гарантирует диапазон $0 < p < 1/2$, в котором выигрыш точно обеспечивается. Оценим точнее: $\alpha = 5\%$, минимальное количество рейтинговых разрядов $G_{min} = 7$, $\epsilon_R \approx \left(\frac{1}{0.05\%}\right)^{\frac{1}{2G_{min}}} - 1 = 0.72$, тогда условие (4.4) – это $p < 81\%$. Т.е. объединение избыточных разрядов даст эффект более устойчивой валидации во всем практическом диапазоне риск-профиля портфеля.

Легко оценить и уровень поправки $\varepsilon$ при $p = 0$ (пренебрегая $\alpha^2$ по сравнению с $\alpha$)

$$\varepsilon \cong \frac{\alpha}{\left(\frac{z_{\alpha/2}}{\epsilon_R} + \frac{z_\alpha}{2}\right)\varphi(z_\alpha)}$$

При условиях выше, поправка получается на уровне $\varepsilon \cong 13.7\%$.

В следующем пункте будут проведены численные расчеты экономии требований к капиталу для портфелей, обладающих риск-профилями, предложенными выше, смотреть Рис. 3, после перехода на различимую РШ за счет более устойчивой положительной оценки валидации, допускающей снижение калибровочной PD.

### 4.2 Численное моделирование экономии требований к капиталу для трех риск-профилей после перехода на различимую рейтинговую шкалу

Имеется три заданных портфеля относительных концентраций $n_i^0, i = 1 \ldots G^0, \sum_{i=1}^{G^0} n_i^0 = 1$ по оригинальным разрядам, у которых задано истинное PD $p_i^0, i = 1 \ldots G^0$, совпадающее в среднем с частотой дефолтов (сглаженное методом (3.1), (3.2)). Портфели: Эксперт-РА, $G^0 = 18$ грейдов; Fitch, $G^0 = 17$ грейдов; ICR, $G^0 = 21$ грейдов. Границы PD грейдов (разрядов РШ) определены формулой (3.3), распределение $F(p)$ строится методом (3.4). Задается ограниченная размерность наблюдений, например $N = 10000$, одинаковая для трех портфелей.

Далее используется метод проектирования статистически различимой рейтинговой шкалы восходящим каскадом (2.3) для заданного количества $N$ наблюдений, который дает новые границы РШ, где по известному распределению $F(p)$ определяются новые относительные концентрации $n_j^D, j = 1 \ldots G^D$ и вероятности дефолта в различимых разрядах $p_j^D, j = 1 \ldots G^D$, которые также истинны в силу построения. Новые размерности $G^D$ для 10000 наблюдений следующие: портфель Эксперт-РА, $G^D = 8$ грейдов; портфель Fitch, $G^D = 8$ грейдов; портфель ICR, $G^D = 9$ грейдов.

Последовательность случайных симуляций для каждого портфеля и двух вариантов РШ (избыточная $G^0$ и различимая $G^D$) следующая:

- Шаг №1. Симулируются случайные концентрации в каждом разряде $N_i^0, N_j^D, N = \sum_{i=1}^{G^0} N_i^0 = \sum_{j=1}^{G^D} N_j^D$ на основе вероятностей их образования $n_i^0, n_j^D$;
- Шаг №2. В каждом разряде $i, j$ симулируется случайное количество дефолтов $ND_i^0, ND_j^D$ из $N_i^0, N_j^D$ на основе вероятностей $p_i^0, p_j^D$ для соответствующих трех РШ;
- Шаг №3. Определяется, нарушен ли критерий Вальда (2.1) в каждом разряде с учетом не асимптотической формулировки (сноска (2.1), для чистоты эксперимента) для заданного уровня значимости $\alpha = 5\%$. Где в качестве PD используются $p_i^0, p_j^D$, размерность наблюдений $N_i^0, N_j^D$, количество наблюдаемых дефолтов $ND_i^0, ND_j^D$ соответственно;

- Шаг №4. Задается критерий критической зоны для РШ: если нарушение для $C = 5$ разрядов и более, то это «красный»; $3 \leq C < 5$ – это «желтый»;
- Шаг №5. Проводится достаточное число $M$ итераций шагов №1-4 (автором использовалось $M = 10000$)

На выходе получаются частоты нарушений критической зоны для двух вариантов РШ – избыточной и различимой.

Однако нам не интересна точечная оценка частот нарушений для заданных $p_i^0, p_j^D$, тем более она близка к нулю в обеих шкалах для $C \geq 5$. Интересна оценка на шаге №3 для критерия «провала теста» в грейде, когда $p_i^0(\varepsilon) = p_i^0 \cdot (1-\varepsilon)$, $p_j^D(\varepsilon) = p_j^D \cdot (1-\varepsilon)$. При этом сами частоты дефолтов порождаются не преобразованными $p_i^0(0) = p_i^0$, $p_j^D(0) = p_j^D$. Теоретические предпосылки, рассмотренные в предыдущем пункте, указывали, что различимая РШ будет иметь преимущество в устойчивости положительного тестирования относительно избыточной РШ.

Для подготовки скалярного аргумента, относительно которого проходят измерения, разумно обратится к метрике, наиболее важной для банка, а именно – требования к капиталу. Для IRB подхода требования к капиталу задаются стандартной формулой, рекомендованной Базельским комитетом (Basel Committee on Banking Supervision, 2023b)[13], которая для LGD=1, задана в виде

$$CR(PD) = EAD \cdot \left( \Phi\left( \frac{\Phi^{-1}(PD) + \sqrt{R} \cdot \Phi^{-1}(0.999)}{\sqrt{1-R}} \right) - PD \right)$$

где $R$ корреляционный коэффициент, заданный в диапазоне от 12% до 24% для корпоративного сегмента. Для целей настоящего исследования бралась средняя величина крупного бизнеса $R = 20\%$. Предполагая гомогенность исследуемых портфелей, положим $EAD = 1$, тогда абсолютные требования к капиталу[14] вычисляются по формуле $CR^0 = \sum_{i=1}^{G^0} n_i^0 \cdot CR(p_i^0)$, аналогично и для различимой РШ $CR^D = \sum_{i=1}^{G^D} n_i^D \cdot CR(p_i^D)$. Очевидно, что переход к различимой РШ увеличивает гранулированность, поэтому разумно ожидать, что $CR^D > CR^0$, что может поставить под сомнение усилия по экономии капитала при переходе на различимую РШ. Однако, расчеты показывают, что это увеличение ничтожно, в долях капитала $\frac{CR^D - CR^0}{CR^0} \cong 0.2 - 0.3\%$, поэтому этим различием можно пренебречь. В качестве аргумента обоснованно взять $VAR\%^0(\varepsilon) = \frac{\sum_{i=1}^{G^0} n_i^0 \cdot CR(p_i^0(\varepsilon))}{CR^0}$, $VAR\%^0(0) = 1$, при нулевом сдвиге $\varepsilon$ капитал остается 100%, при не нулевом – уменьшается.

---

[13] CRE 31.5

[14] В процентах от экспозиции портфеля

Для демонстрации эффекта перехода на различимую РШ по оси абсцисс откладывается $VAR\%^0(\varepsilon)$, по оси ординат – доля (%) провальных валидаций двух РШ $BAD\%^0(\varepsilon), BAD\%^D(\varepsilon)$ относительно критерия C на Шаге №4 симуляций.

На Рис. 8 представлен эффект снижения капитала относительно процентной доли провала валидации при «красном критерии» $C \geq 5$ для портфеля ICR.

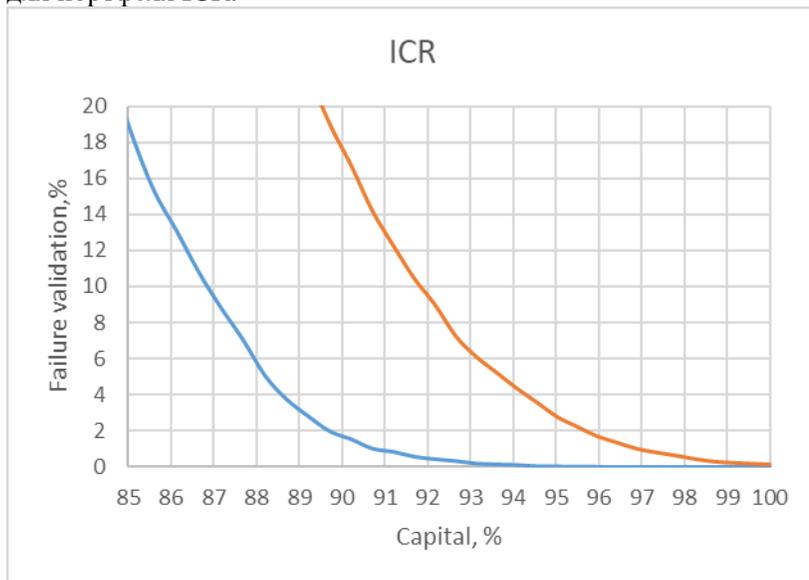

**РИСУНОК 8.** Зависимость доли провальных валидаций относительно снижения требований к капиталу для портфеля ICR при «красном» критерии $C \geq 5$ и $N = 10000$ наблюдений для избыточной РШ (коричневый) и различимой РШ (синий).

Из Рис.8 видно, что эффект начинается с незначительных долей провальных валидаций и устойчив по мере их увеличения. Поэтому имеет смысл зафиксировать допустимую долю «провалов» на уровне 1% и применить анализ разницы экономии капитала для всех трех исследуемых портфелей для разных размерностей наблюдений, но неизменной различимой РШ, построенной на 10000 наблюдений. На Рис. 9 представлены результаты экономии капитала (относительно начального значения, принимаемого за 100%) как функция от размерности наблюдений для трех исследуемых портфелей.

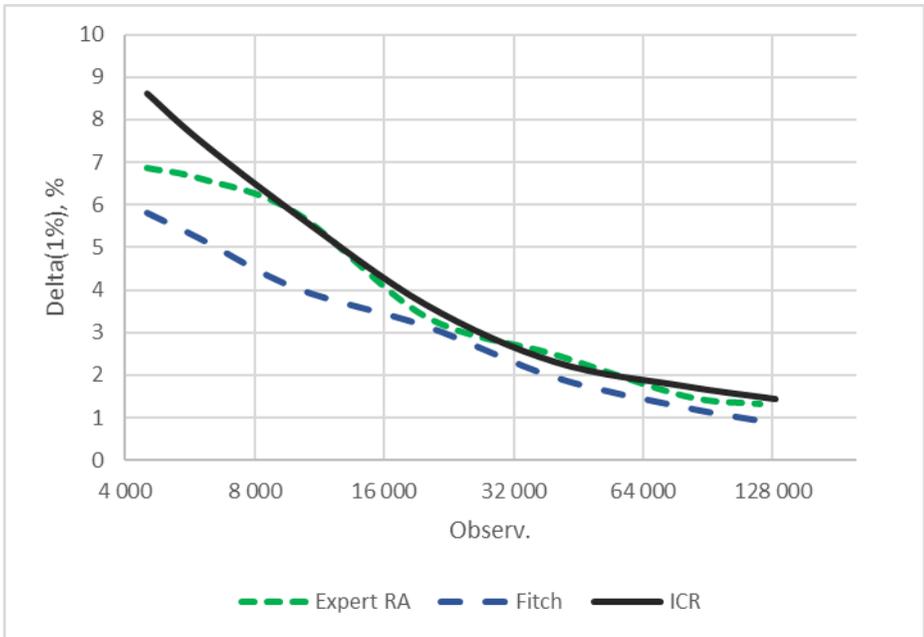

**РИСУНОК 9.** Зависимость эффекта экономии капитала при переходе на различимую постоянную РШ, построенную на 10000 наблюдений, для трех исследуемых портфелей как функция от размерности наблюдений при валидации, на которой вероятность «провала» равна один процент.

Из Рис.9. видно, что что эффект экономии капитала при переходе на различимую при 10000 наблюдений РШ нивелируется по мере увеличения размерности наблюдений. Поэтому, для массового кредитования (физлица, МСБ) представленный подход эффекта экономии даст немного.

Следующий вопрос исследования, сохраниться ли эффект, если изменить критический уровень критерия, например на «желтый», $C \geq 3$. Ответ утвердительный, Рис. 10.

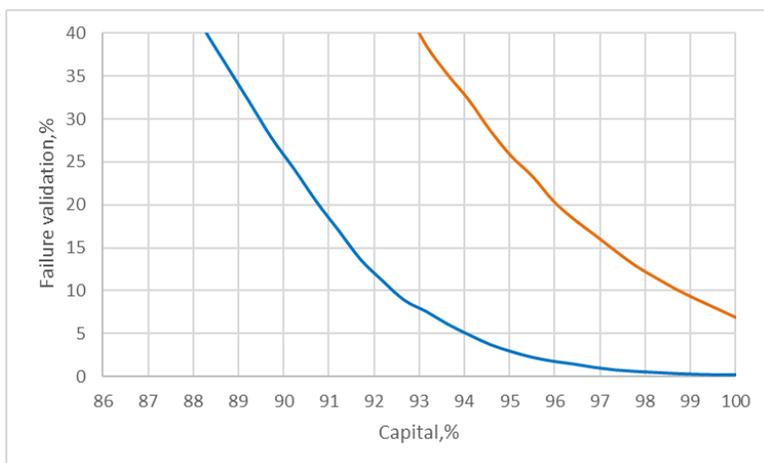

**РИСУНОК 10.** Зависимость доли провальных валидаций относительно снижения требований к капиталу при «желтом» критерии $C \geq 3$ и $N = 10000$ наблюдений для избыточной РШ (коричневый) и различимой РШ (синий).

Отличие Рис.10 («желтый» критерий) от Рис.8 («красный» критерий) только в том, что избыточная РШ даже без снижения PD (экономии капитала) уже порождает заметную вероятность «провала валидации». Точное ее значение 6.8%, если допускать ту же вероятность «провала валидации» для различимой РШ, то можно сэкономить 6.5% процентов капитала.

## 5 ЗАКЛЮЧЕНИЕ

В представленной работе предложен механизм оптимизации РШ и показано, что проектирование шкалы – это задача, которую нужно решать с учетом статистики наблюдений, на которые ориентирована РШ. Предполагается, что шкала разрабатывается на основе прямого маппинга. В работе предложены формулы для каскада определения границ PD рейтинговых разрядов, представлены графики зависимости размерности полученной РШ от количества наблюдений «на руках» у разработчика РШ. Основная идея – это требование, чтобы каждый разряд был различим с точки зрения критерия Вальда биномиального теста. Снижение размерности РШ и установка границ, оптимальных с точки зрения различимости, приводит к более надежной валидации РШ, которая позволяет снизить калибровочное PD без заметного увеличения риска эту валидацию не пройти. Этот эффект получил теоретическое обоснование и продемонстрирован на трех практических РШ - двух рейтинговых агентств и одного банка. Эффект приводит к экономии требований к капиталу на уровне 6-8% относительно первоначальной величины, рассчитанной на избыточной РШ для ограниченной размерности наблюдений. При существенной размерности наблюдений, более 100-200 тыс., эффект нивелируется. Предложенный метод проектирования РШ наиболее актуален для корпоративного сегмента кредитного портфеля, на который, как правило, расходуется большая часть капитала банка. Подразумевается, что оценка требований к капиталу проводится по продвинутой формуле IRB подхода, рекомендованной Базельским комитетом. Внедрение различимой РШ должно состояться до первичной валидации рейтинговой модели корпоративного сегмента, поскольку выигрыш можно получить только если настраивать калибровку по уже внедренной РШ, а не наоборот.

## DECLARATION OF INTEREST

The authors report no conflicts of interest. The authors alone are responsible for the content and writing of the paper.

26   M. Pomazanov